\documentclass[12pt]{article}

\usepackage{amsthm,amsmath,amsfonts,braket,algorithm,graphicx, braket}
\usepackage{fullpage,authblk}

\theoremstyle{definition} 
\theoremstyle{definition} 
\newtheorem {theorem} {Theorem}

\newtheorem {lemma} {Lemma}

\newcommand{\kb}[1]{\mathbf{\left[#1\right]}}

\newcommand{\al}{\mathcal{A}}

\newcommand{\trd}[1]{\left|\left| #1 \right| \right|}

\newcommand{\st}{\text{ } : \text{ }}

\newcommand{\Hmin}{H_\infty}

\newcommand{\leak}{\texttt{leak}_{EC}}

\floatname{algorithm}{Protocol}

\begin{document}
\title{High-Dimensional Quantum Conference Key Agreement}
\date{}
\author[1,2]{Omar Amer}
\author[1]{Walter O. Krawec\footnote{Email: \texttt{walter.krawec@uconn.edu}}}
\affil[1]{\small{Department of Computer Science and Engineering}\\\small{University of Connecticut}\\\small{Storrs, CT 06269 USA}}
\affil[2]{\small{Future Lab for Applied Research and Engineering}\\\small{JPMorgan Chase Bank, N.A}\\\small{New York, NY 10017 USA}}

\maketitle
	
\begin{abstract}
Quantum Conference Key Agreement (QCKA) protocols are designed to allow multiple parties to agree on a shared secret key, secure against computationally unbounded adversaries.  In this paper, we consider a high-dimensional QCKA protocol and prove its information theoretic security against arbitrary, general, attacks in the finite-key scenario.  Our proof technique may be useful for other high-dimensional multi-party quantum cryptographic protocols.  Finally, we evaluate the protocol in a variety of settings, showing that high-dimensional states can greatly benefit QCKA protocols.
\end{abstract}

\section{Introduction}

Quantum key distribution (QKD) allows for the establishment of a shared secret key between two parties, Alice and Bob, secure against computationally unbounded adversaries (whom we refer to as Eve).  Progress in these protocols has rapidly advanced, leading to both a rich theory along with practical commercial systems \cite{qkd-survey-scarani,qkd-survey-pirandola,amer2021introduction}.  Quantum conference key agreement (QCKA) protocols are designed to allow multiple parties to establish a common, shared, secret key secure against computationally unbounded adversaries.  Starting from early work in this field \cite{group-key-first,group-key-newer}, QCKA protocols have advanced substantially with new protocols and security proofs \cite{grasselli2019conference,wu2016continuous,ottaviani2019modular}; it is also experimentally feasible \cite{proietti2021experimental}.  Interestingly, it has been shown that there are some scenarios where such multiparty protocols hold an advantage over the naive use of multiple two-party protocols run in parallel \cite{group-key-newer}.  For a recent survey on quantum conference key agreement protocols and the state of the art in security proofs, the reader is referred to \cite{murta2020quantum}.


High-dimensional quantum cryptography has been shown to exhibit numerous advantages over qubit-based protocols, especially in two-party QKD  \cite{bechmann2000quantum,chau2005unconditionally,sheridan2010security,sasaki2014practical,chau2015quantum,vlachou2018quantum,cerf2002security,nikolopoulos2005security,iqbal2020high,nikolopoulos2006error,yin2018improved,doda2021quantum}.  Encouraged by this, it is worth investigating whether high-dimensional states can benefit QCKA.  To our knowledge, only one high-dimensional QCKA protocol exists which was introduced in \cite{pivoluska2018layered}, however no rigorous finite key security analysis exists for it (instead, \cite{pivoluska2018layered} developed layered QKD protocols and was not concerned with the explicit finite-key analysis of this particular QCKA protocol - in fact, our analysis done in this paper may be useful in proving security of those other protocols introduced in \cite{pivoluska2018layered}, though we leave that as interesting future work).

In this work, we consider a high-dimensional QCKA protocol and prove its security against arbitrary, general attacks in the finite key setting.  The protocol we analyze is an extension of the qubit-based protocol from \cite{finite-ghz-bb84} to higher dimensions and also a specific instance of a protocol introduced in \cite{pivoluska2018layered}.  For the security proof, we utilize the quantum sampling framework introduced by Bouman and Fehr in \cite{sample}, along with proof techniques we developed in \cite{krawec2019quantum} to derive sampling-based entropic uncertainty relations.  Our proof, though using these two frameworks as a foundation, introduces several new methods which may also be useful when analyzing other quantum cryptographic protocols, both those involving two users and those for multi-users, especially in higher dimensions.

Finally, we evaluate the performance of this protocol in a variety of scenarios, showing some very interesting behavior and shedding new light on the benefits of high-dimensional quantum states.  In particular, we show that, as the dimension of the quantum signal increases, the noise tolerance also increases.  Interestingly, the key-rate also increases beyond what would be possible by simply running multiple, lower-dimensional, protocols in parallel.  This shows that high-dimensional states can greatly benefit QCKA protocols.  Our contributions in this work are not only in developing a security proof for a high dimensional QCKA protocol, but also in showing even more benefits to high-dimensional quantum states when applied to quantum cryptography.  Our methods may also spur future research in this area, as our proof techniques may be highly adaptable to other scenarios.


\subsection{Notation and Definitions}

We begin with some notation and definitions that we will use in this work.  Let $d \in \mathbb{N}$, then we write $\al_d$ to be a $d$-character alphabet with a distinguished $0$ element.  Given a word $q \in \al_d^n$, and a subset $t \subset \{1, \cdots, n\}$, we write $q_t$ to mean the substring of $q$ indexed by $t$; we use $q_{-t}$ to mean the substring of $q$ indexed by the complement of $t$.  We write $w(q)$ to be the relative Hamming weight of $q$, namely $w(q) = \frac{|\{i \st q_i \ne 0\}|}{n}$ - that is the number of characters in $q$ that are not zero, divided by the length of $q$.  Given two words $x, y$ in this alphabet, we write $xy$ to mean the concatenation of $x$ and $y$.  Finally, given $a,b$, numbers between $0$ and $d-1$, we write $a +_d b$ to mean the addition of $a$ and $b$ modulo $d$.

We use $\mathcal{H}_d$ to mean a Hilbert space of dimension $d$.  The standard computational basis will be denoted $Z = \{\ket{0}, \ket{1}, \cdots, \ket{d-1}\}$.  If we are referring to an alternative basis we will write the basis label as a superscript.  One important basis we will use is the Fourier basis consisting of elements $\mathcal{F} = \{\ket{0}^\mathcal{F}, \cdots, \ket{d-1}^{\mathcal{F}}\}$, where:
\[
\ket{j}^\mathcal{F} = \frac{1}{\sqrt{d}}\sum_k \exp(2\pi i j k /d) \ket{k}.
\]
If given a word $q \in \al_d^n$, we write $\ket{q}$ to mean $\ket{q_1}\otimes\cdots\otimes \ket{q_n}$.  Similarly, we write $\ket{q}^\mathcal{F}$ to mean $\ket{q_1}^\mathcal{F}\otimes\cdots\otimes\ket{q_n}^{\mathcal{F}}$.  Note that if there is no superscript, then $\ket{q}$ is assumed to be the computational $Z$ basis.  Finally, given pure state $\ket{\psi}$, we write $\kb{\psi}$ to mean $\ket{\psi}\bra{\psi}$.

A density operator is a positive semi-definite Hermitian operator of unit trace acting on some Hilbert space.  If $\rho_{AE}$ acts on Hilbert space $\mathcal{H}_A\otimes\mathcal{H}_E$, then we write $\rho_A$ to mean the operator resulting from tracing out the $E$ system, namely $\rho_A = tr_E\rho_{AE}$.  Similarly for other, or multiple, systems.

The Shannon entropy of a random variable $X$ is denoted $H(X)$.  The $d$-ary entropy function is denoted $H_d(x)$, for $x \in [0,1]$, and is defined to be:
\[
H_d(x) = x\log_d(d-1) - x\log_d x - (1-x)\log_d (1-x).
\]
Note that when $d=2$ this is simply the binary Shannon entropy.  Given density operator $\rho_{AE}$, the conditional \emph{quantum min entropy} is defined to be \cite{renner2008security}:
\begin{equation}
\Hmin(A|E)_\rho = \sup_{\sigma_E}\max\{\lambda\in\mathbb{R} \st 2^{-\lambda}I_A\otimes\sigma_E - \rho_{AE} \ge 0\},
\end{equation}
where the supremum is over all density operators acting on the $E$ system.  If $\rho = \kb{\psi}$ is a pure state, then we often write $\Hmin(A|E)_\psi$.  Given $\rho_{AE}$, we write $\Hmin(A_Z|E)_\rho$ to mean the min entropy of the resulting state following a measurement of the $A$ register in the $Z$ basis.

There are many important properties of quantum min entropy we will use.  In particular, if the $E$ system is trivial or independent of the $A$ system, then $\Hmin(A)_\rho = -\log_2\max\lambda$, where the maximum is over all eigenvalues $\lambda$ of $\rho_A$.  Given a state $\rho_{AEC} = \sum_{c=0}^Mp_c\rho_{AE}^{(c)}\otimes \kb{c}$ (i.e., the $C$ register is classical), then:
\begin{equation}\label{eq:qc-state}
\Hmin(A|EC)_\rho \ge \min_c\Hmin(A|E)_{\rho^{(c)}}.
\end{equation}

An important result proven in \cite{sample}, based on a lemma in \cite{renner2008security}, is the following which allows one to compute the min entropy of a superposition state based on the min entropy of a suitable mixture state:
\begin{lemma}\label{lemma:superposition}
  (From \cite{sample}): Let $Z$ and $X$ be two orthonormal bases of $\mathcal{H}_d$.  Then for any pure state $\ket{\psi}_{AE} = \sum_{i\in J}\alpha\ket{i}^X\otimes\ket{E_i}$, with $J \subset \al_d^N$, it holds that:
  \[
  \Hmin(A_Z|E)_\psi \ge \Hmin(A_Z|E)_\rho - \log_2|J|,
  \]
  where $\rho_{AE} = \sum_{i\in J}|\alpha_i|^2\kb{i}^X\otimes\kb{E_i}$, and where the entropies above are computed on the state following a $Z$ basis measurement.
\end{lemma}
Quantum min-entropy is a vital resource in QKD security.  Indeed, given a classical-quantum state $\rho_{AE}$, then the amount of uniform independent randomness that may be extracted from the $A$ register after a privacy amplification process is a function of conditional min entropy.  In particular, let $\sigma_{KE}$ be the resulting state after privacy amplification (a process of hashing the $A$ register to a size of $\ell$ bits using a randomly chosen two-universal hash function), then it was shown in \cite{renner2008security} that:
\begin{equation}\label{eq:PA}
\trd{\sigma_{KE} - I/2^\ell \otimes \sigma_E} \le 2^{-\frac{1}{2}(\Hmin(A|E)_\rho - \ell)}.
\end{equation}

In our security proof, we will utilize a quantum sampling framework originally introduced in 2010 by Bouman and Fehr \cite{sample} and used by us recently to prove novel sampling-based entropic uncertainty relations \cite{krawec2019quantum,krawec2020new} and proofs of security for high-dimensional BB84 \cite{yao2022quantum}.  We review some of the terminology and results from \cite{sample} here; for more information on these results, the reader is referred to that original reference.

Fix $d \ge 2$ and $N \ge 1$.  A \emph{classical sampling strategy} is a tuple $(P_T, f, g)$ where $P_T$ is a distribution over all subsets of $\{1, \cdots, N\}$ and $f, g: \al_d^* \rightarrow \mathbb{R}$.  Given $q \in \al_d^N$, the strategy will first choose $t$ according to $P_T$; it will then observe $q_t$ and evaluate $f(q_t)$.  This evaluation should be a ``guess'' as to the value of some target function, $g$, evaluated on the \emph{unobserved} portion.  Namely, for a good sampling strategy, with high probability over the choice of subset $t$, it should hold that $f(q_t)$ is $\delta$-close to $g(q_{-t})$ for given $\delta > 0$.

More formally, fix a subset $t$ with $P_T(t) > 0$.  We define the set of ``good'' words $\mathcal{G}_t$ to be:
\begin{equation}\label{eq:good-words}
\mathcal{G}_t = \{q \in \al_d^N \st |f(q_t) - g(q_{-t})| \le \delta\}
\end{equation}
Note that, given $q \in \mathcal{G}_t$, if subset $t$ were to be chosen by the sampling strategy, it is guaranteed that the strategy will succeed (the guess will be $\delta$-close to the target value).  The \emph{error probability} of the sampling strategy, then, is:
\[
\epsilon^{cl} = \max_{q\in\al_d^N}Pr\left(q \not \in \mathcal{G}_t\right),
\]
where the probability is over all subsets chosen according to $P_T$.  One sampling strategy we will need later is summarized in the following lemma:
\begin{lemma}\label{lemma:sample}
(From \cite{sample}): Let $\delta > 0$ and $m \le N/2$.  Define $P_T$ to be the uniform distribution over all subsets of $\{1, \cdots, N\}$ of size $m$.  Define $f(x) = g(x) = w(x)$.  Then:
\[
\epsilon^{cl} \le 2\exp\left(\frac{-\delta^2m N}{N+2}\right).
\]
\end{lemma}

These definitions may be promoted to the quantum case.  Fixing a sampling strategy and a $d$-dimensional basis $\mathcal{B}$, we define $\text{span}(\mathcal{G}_t) = \text{span}(\ket{q}^\mathcal{B} \st q \in \mathcal{G}_t)$.  Note that, for any $\ket{\psi} \in \text{span}(\mathcal{G}_t)\otimes\mathcal{H}_E$, if a measurement in the $\mathcal{B}$ basis were made on those qudit systems indexed by $t$ resulting in outcome $q \in \al_d^{|t|}$, it would hold that the collapsed post-measured state must be of the form:
\[
\ket{\psi_t^q} = \sum_{x\in J_q}\alpha_x\ket{x}^\mathcal{B}\otimes\ket{E_x},
\]
where $J_q = \{x \in \al_d^{N-|t|} \st |f(q) - g(x)| \le \delta\}$.

The main result from \cite{sample} may then be stated as follows:
\begin{theorem}\label{thm:sample}
(From \cite{sample} though reworded for our application in this work): Let $(P_T, f, g)$ be a classical sampling strategy with error probability $\epsilon^{cl}$ for a given $\delta > 0$ and let $\ket{\psi}_{AE}$ be a quantum state where the $A$ register lives in a Hilbert space of dimension $d^N$.  Then, there exist ideal states $\ket{\phi^t} \in \text{span}(\mathcal{G}_t)\otimes\mathcal{H}_E$ (with respect to some given, fixed, $d$-dimensional basis $\mathcal{B}$) such that:
\begin{equation}\label{eq:ideal-real}
\frac{1}{2}\trd{\sum_tP_T(t)\kb{t}\otimes\left(\kb{\psi} - \kb{\phi^t}\right)} \le \sqrt{\epsilon_{cl}}.
\end{equation}
where the above summation is over all subsets $t\subset\{1, \cdots, N\}$.
\end{theorem}
Note that the above is a slight rewording of the main result from \cite{sample}.  For a proof that Theorem \ref{thm:sample} follows from the main result in \cite{sample}, the reader is referred to \cite{yao2022quantum}.

\section{Protocol}

The protocol we consider is a high-dimensional variant of the QCKA agreement protocol originally introduced and analyzed in \cite{finite-ghz-bb84}.  It is also a specific instance of a protocol introduced for a layered QKD system in \cite{pivoluska2018layered} (though without a complete proof of security).  We assume there are $p$ Bob's and one Alice all of whom wish to agree on a shared secret group key.  The protocol begins by having Alice prepare the following high-dimensional GHZ state:
\[
\ket{\psi_0} = \frac{1}{\sqrt{d}}\sum_{a=0}^{d-1} \ket{a, \cdots, a}_{AB_1\cdots, B_p}.
\]
Above, $d$ is the dimension of a single system ($d=2$ in the protocol analyzed in \cite{finite-ghz-bb84}).  The $B_i$ system is sent to the $i$'th Bob while Alice retains the $A$ register.  Randomly, Alice and the $p$ Bob's will measure their registers in the Fourier basis $\mathcal{F}$ resulting in outcome $q_{AB_1\cdots B_p} \in \al_d^{p+1}$.  Otherwise, if  Alice and the $p$ Bob's choose not to measure in the Fourier basis, they will measure in the computational basis, the result of which will be used to add $\log_2 d$ bits to their raw key.  Note that the choice of whether to measure in the Fourier basis or the computational $Z$ basis may be made randomly by all parties (discarding events when choices are not consistent) or by using a pre-shared secret key (as was done in \cite{finite-ghz-bb84}).  The above process is repeated for a freshly prepared and sent $\ket{\psi_0}$ until a raw key of sufficient length has been established.  Note that, in the original qubit-based version introduced in \cite{finite-ghz-bb84}, the Hadamard $X$ basis was used instead of explicitly the Fourier basis - however both are equivalent in dimension two; in higher dimensions, we use the Fourier basis for this test measurement.  This protocol here, generalizes the one from \cite{finite-ghz-bb84} to higher dimensions and when $d=2$ they are equivalent protocols.

Interestingly, unlike standard BB84 \cite{QKD-BB84} (or, rather, the entanglement based version E91 \cite{QKD-E91}), measuring in an alternative, non computational, basis cannot lead to a correlated secret key digit as the results will not be identical for all parties.  However, the Fourier basis measurement can be used to test for errors, leaving the $Z$ basis measurement alone for key distillation.  Note that, if there is no noise in the channel, it should hold that whenever parties measure in the $\mathcal{F}$ basis, the results should sum to $0$ modulo $d$, namely: $q_A +_d q_{B_1} +_d\cdots +_d q_{B_p} = 0$; any non-zero sum will be considered noise and factored into our key-rate analysis.  That this is true is easy to see.  Indeed, converting $\ket{\psi_0}$ to the Fourier basis yields:
\[
\ket{\psi_0} = \frac{1}{\sqrt{d}}\sum_{a=0}^{d-1}\left(\sum_{j_0,\cdots, j_p\in\al_d}\frac{1}{\sqrt{d^{p+1}}}\exp(-2\pi i (j_0+\cdots+j_p)a/d)\ket{j_0,\cdots,j_p}^\mathcal{F}\right)
\]
Now, if $j_0+\cdots + j_p = \lambda \cdot d$ for some $\lambda \in \mathbb{Z}$, then the probability of observing that particular $\ket{j_0,\cdots,j_p}^\mathcal{F}$ is simply:
\[
\frac{1}{d^{p+2}}\trd{\sum_{a=0}^{d-1}\exp(-2\pi i a)}^2 = \frac{1}{d^{p+2}}\times d^2 = \frac{1}{d^p}.
\]
Since there are exactly $d^p$ such $j_0,\cdots, j_p \in \al_d$ and their sum, modulo $d$ is zero, it follows that the only observable values in the Fourier basis must sum to a number divisible by the dimension $d$.  This proves the protocol is correct - namely, if the source is ideal, parties will distill a correlated key and not abort since their test measurement will result in the prescribed all-zero string.

Following the establishment of the raw key, Alice and the $p$ Bob's will run a pair-wise error correction protocol followed by a standard privacy amplification protocol.  Following error correction, but before privacy amplification, Alice will choose a random two-universal hash function $f$, the output size of which we take to be $\log_2\frac{1}{\epsilon_{EC}}$-bits (for user-specified $\epsilon_{EC}$), and broadcast $f(A)$, where $A$ is her raw key.  Each Bob will locally compare the result of running their version of the raw key through this hash function and if the digest doesn't match, all parties abort.  This ensures that, except with probability at most $\epsilon_{EC}$, parties can be assured that error correction has succeeded. This, of course, leaks an additional $\log_2\frac{1}{\epsilon_{EC}}$ bits which must be deducted from the final secret key size.  We will comment more on error correction later when evaluating our key-rate bound.


\section{Security Proof}


To prove security of this protocol, we analyze the security of an equivalent entanglement based version.  Here, instead of having Alice prepare and send a quantum state, we allow Eve the ability to create any arbitrary initial state, sending part to Alice and the other parts to the $p$ Bob's while also potentially maintaining a private entangled ancilla.  Clearly security in this case will imply security of the prepare-and-measure version discussed in the previous section.  We also use as a foundation, a proof methodology we introduced in \cite{krawec2019quantum}, though making several modifications for the multi-party protocol being analyzed here.  Our proof of security, at a high level, proceeds in three steps: first we define an analyze an appropriate classical sampling strategy allowing us to use Theorem \ref{thm:sample}; second, we analyze the ideal states produced by that Theorem; and third, finally, we promote that ideal-case analysis to the real state.

\textbf{Entanglement Based Protocol -} Let $\ket{\psi}\in \mathcal{H}_A\otimes\mathcal{H}_{B_1}\otimes\cdots\otimes\mathcal{H}_{B_p}\otimes\mathcal{H}_E$ be the state Eve prepares where each $\mathcal{H}_A \cong \mathcal{H}_{B_i} \cong \mathcal{H}_d^{\otimes N}$.  Here $N$ is the user-specified number of rounds used by the protocol and is a parameter users may optimize.  Ideally $\ket{\psi} = \ket{\psi_0}^{\otimes N}$.  At this point, the users choose a random subset $t \subset \{1, 2, \cdots, N\}$ of size $m < N/2$ for sampling.  This can be done by having Alice choose the subset and sending it to the Bob's (the option we assume here) or by using a small pre-shared key (the option used in \cite{finite-ghz-bb84}).  Each party will measure their respective $d$ dimensional signals, indexed by $t$, in the $d$-dimensional Fourier basis, $\mathcal{F}$, resulting in outcome ${q} = q_Aq_{B_1}\cdots q_{B_p} \in \al_d^{m(p+1)}$.  Here, each $q_A, q_{B_1}, \cdots, q_{B_p}$ is an $m$ character string which we may enumerate as $q_A = q_A^1\cdots q_A^m$ and $q_{B_i} = q_{B_i}^1\cdots q_{B_i}^m$.

Let $s_i({q}) = q_A^i +_d q_{B_1}^i +_d \cdots +_d q_{B_p}^i$.  That is, $s_i$ is the sum, modulo the dimension $d$, of all user measurement outcomes for signal $i$.  Also, define $s(q) = s_1(q)\cdots s_m(q) \in \al_{d}^m$.  If the source $E$ were honest, it should be that $w(s(q)) = 0$ since this will be the case in the event Eve prepared copies of $\frac{1}{\sqrt{d}}\sum_{a=0}^{d-1}\ket{a, a, \cdots, a}_{AB_1\cdots B_p}$ as discussed earlier.

\textbf{Step 1: Classical Sample Strategy Analysis -} We now wish to use Theorem \ref{thm:sample} to analyze the security of this protocol.  To do so, we require a suitable classical sampling strategy which corresponds to the sampling done by the actual protocol, and a bound on its error probability.  Consider the following classical sampling strategy: given a word $q = q^0q^1q^2\cdots q^p \in \al_d^{(p+1)\cdot N}$ (i.e., each $q^j \in \al_d^N$), then first choose a subset $t \subset \{1, \cdots, N\}$ of size $m \le N/2$ and observe $q_t = q_t^0q^1_tq^2_t\cdots q^p_t$ (namely, one observes the $t$ portion of each of the $p+1$ strings).  From this, compute $f(q_t) = w(s(q_t))$ to estimate the value of $g(q_{-t}) = w(s(q_{-t}))$.  Putting this into the notation introduced earlier, we have the set of ``good'' words (see Equation \ref{eq:good-words}) as:
\[
\mathcal{G}_t = \{q \in \al_d^{(p+1)\cdot N} \st |w(s(q_t)) - w(s(q_{-t}))| \le \delta\}.
\]
This is exactly the sampling strategy we wish to use in our QCKA protocol.  Users will observe a value based on their measurement in the Fourier basis, in particular, they observe the number of outcomes that do not sum to $0$ modulo $d$.  We wish to argue that the remaining, unmeasured portion, satisfies a similar restriction in the $\mathcal{F}$ basis, thus placing a constraint on the form of the state Eve prepared, needed to compute the min entropy later.  In order to use Theorem \ref{thm:sample}, needed to construct suitable ideal quantum states, we require a bound on the error probability of this classical sampling strategy.  In particular, we require:
\[
\epsilon^{cl} = \max_{q\in\al_d^{(p+1)N}}Pr\left(q \not\in \mathcal{G}_t\right).
\]

We claim:
\begin{equation}
\epsilon^{cl} \le 2\exp\left(\frac{-\delta^2m N}{N+2}\right).
\end{equation}
Let $\widetilde{\mathcal{G}}_t = \{q \in \al_d^N \st |w(q_t) - w(q_{-t})| \le \delta\}$.  Note that, by Lemma \ref{lemma:sample}, it holds that:
\[
\tilde{\epsilon}^{cl} = \max_{\tilde{q}\in\al_d^N}Pr(\tilde{q}\not\in\widetilde{\mathcal{G}}_t) \le 2\exp\left(\frac{-\delta^2m N}{N+2}\right).
\]
Pick $q \in \al_d^{(p+1)N}$ and let $\tilde{q} = s(q)$.  Then, it is clear that if $q \not \in \mathcal{G}_t$ then $\tilde{q} \not \in \widetilde{G}_t$ for any subset $t$.  Thus for every $q\in\al_d^{(p+1)N}$, it holds that $Pr(q\not\in \mathcal{G}_t) \le Pr(\tilde{q}\not\in\widetilde{\mathcal{G}}_t)$ from which the claim follows.

\textbf{Step 2: Ideal State Analysis -} We now return to the security analysis of the protocol.  Let $\epsilon > 0 $ be given (it will, as we discuss later, determine the security level of the secret key).  From Theorem \ref{thm:sample}, using the above sampling strategy with respect to the Fourier basis, there exists an ideal state of the form $\frac{1}{T}\sum_t\kb{t} \otimes \kb{\phi^t}$ where $T = {N \choose m}$ and:
\begin{equation}
  \ket{\phi^t} \in \text{span}\{\ket{q}^\mathcal{F} \st q \in \al_d^{(p+1)N} \text{ and } |w(s(q_t)) - w(s(q_{-t}))|\le\delta\}.
\end{equation}
If we set
\begin{equation}
\delta = \sqrt{\frac{(m+n+2)\ln(2/\epsilon^2)}{m(m+n)}}.
\end{equation}
then, we have that the real and ideal states are $\epsilon$-close in trace distance (on average over the subset choice as shown in Equation \ref{eq:ideal-real}) with the real-state being $\frac{1}{T}\sum_{t}\kb{t}\otimes\kb{\psi}$.

We first analyze the ideal case and then use this analysis to argue about security of the actual given input state from Eve.  In the ideal case, the event of choosing subset $t$, measuring those systems in the Fourier basis and observing outcome ${q}\in\al_d^{(p+1)m}$, causes the ideal state to collapse to:
\begin{equation}\label{eq:ideal-post-measure}
  \ket{\phi_q^t} = \sum_{x\in J_q}\alpha_x\ket{x}^\mathcal{F}\otimes\ket{E_x},
\end{equation}
where:
\begin{align}
  J_q &= \{x_Ax_{B_1}\cdots x_{B_p} \in \al_d^{(p+1)n} \st |w(s(x)) - w(s(q))|\le\delta\}\notag\\\notag\\
  &= \left\{x_A^1\cdots x_A^n x_{B_1}^1\cdots x_{B_1}^n\cdots x_{B_p}^1\cdots x_{B_p}^n \text{ such that }\right.\\
  & \left.|w( [x_A^1 +_d \cdots +_d x_{B_p}^1]\cdots [x_A^n +_d \cdots +_d x_{B_p}^n]) - w(s(q))|\le\delta\right\}. \notag
\end{align}
By manipulating the above state, we may write it in the following form which will be more useful for us in our analysis:
\begin{align}
  \ket{\phi_q^t} \cong \sum_{\substack{
      x_{B_1}^1\cdots x_{B_1}^n = x_{B_1}\in\al_d^n\\
      x_{B_2}^1\cdots x_{B_2}^n = x_{B_2}\in\al_d^n\\
      \vdots\\
      x_{B_p}^1\cdots x_{B_p}^n = x_{B_p}\in\al_d^n
  }}
  \beta_{x}\ket{x}^\mathcal{F}_{B_1\cdots B_p} \otimes \sum_{y \in J(q \st x)} \beta_{y|x}\ket{y}^\mathcal{F}_A\ket{F_{x,y}}_E
\end{align}
where, above, we define $x = x_{B_1}\cdots x_{B_p} \in \al_d^{p\cdot n}$ and we define:
\begin{equation}
  J(q\st x) = \{y\in\al_d^n \st |w(s(yx)) - w(s(q))|\le\delta\}
\end{equation}
Note that some of the $\beta$'s in the above expression may be zero; also note that we permuted the subspaces above to place the $A$ register to the right of the $B$ registers - this was done only to make the algebra in the remainder of the proof easier to follow.

Our goal now is to compute a lower bound on the conditional quantum min entropy following a $Z$ basis measurement on the collapsed ideal state (that is, the entropy in the above state $\ket{\phi_q^t}$, but following Alice's $Z$ basis measurement on her $A$ register).  Tracing out $B$'s system yields:
\begin{equation}
  \sigma_{AE} = \sum_{x\in\al_d^{p\cdot n}}|\beta_{x}|^2\underbrace{P\left(\sum_{y\in J(q \st x)} \beta_{y|x}\ket{y}^\mathcal{F}_A\ket{F_{x,y}}_E\right)}_{\sigma_{AE}^{(x)}},
\end{equation}
where $P(\ket{z}) = \kb{z}$.

From Equation \ref{eq:qc-state}, we have $\Hmin(A_Z|E)_\sigma \ge \min_x\Hmin(A_Z|E)_{\sigma^{(x)}}$.  Fix a particular $x$ and consider the mixed state:
\begin{equation}
  \chi^{(x)}_{AE} = \sum_{y\in J(q\st x)}|\beta_{y|x}|^2\kb{y}_A^\mathcal{F}\otimes\kb{F_{x,y}}_E.
\end{equation}
From Lemma \ref{lemma:superposition}, we have:
\[
\Hmin(A_Z|E)_{\sigma^{(x)}} \ge \Hmin(A_Z|E)_{\chi^{(x)}} - \log_2|J(q\st x)|.
\]
We first compute a bound on the size of $J(q\st x)$.  Let $\mathcal{I} = \{y\in\al_d^n\st |w(y) - w(s(q))|\le\delta\}$.  We claim $|J(q\st x)| \le |\mathcal{I}|$.  Indeed, pick $y \in J(q\st x)$ and let $z = s(yx)$.  Then $z \in \mathcal{I}$.  Furthermore, for any $y,y' \in J(q\st x)$ with $y \ne y'$, it holds that $s(yx) \ne s(y'x)$.  Thus the claim follows.  Now, since $|\mathcal{I}| \le d^{nH_d(w(s(q)) + \delta)}$ by the well known bound on the volume of a Hamming ball, we have an upper-bound on the size of the set $J(q\st x)$ as a function of the observed value $q$. Note that, ideally, $w(s(q)) = 0$ with non-zero values representing error in the channel, and so the size of this set should be ``small'' for low noise levels.  As the noise increases, our entropy bound will decrease (thus ultimately decreasing the overall key-rate as expected).

What remains is to compute $\Hmin(A_Z|E)_{\chi}$.  Following a $Z$ basis measurement on the $A$ register in $\chi$, we are left with the post-measured state:
\begin{equation}
  \chi_{A_ZE} = \sum_y |\beta_{y|x}|^2 \sum_{z\in\al_d^n}p(z|y)\kb{z}_A\kb{F_{x,y}}_E,
\end{equation}
where $p(z|y)$ is the conditional probability of observing outcome $\ket{z}$ given input state $\ket{y}^\mathcal{F}$.  Now, consider the following state where we add an additional, classical, ancilla:
\[
\chi_{A_ZEY} = \sum_y |\beta_{y|x}|^2 \kb{y}_Y\otimes \underbrace{\sum_{z\in\al_d^n}p(z|y)\kb{z}_A\kb{F_{x,y}}_E}_{\chi^{(y)}}.
\]
Then we have $\Hmin(A_Z|E)_{\chi} \ge \Hmin(A_Z|EY)_{\chi} \ge \min_y\Hmin(A_Z|E)_{\chi^{(y)}}$ where we used Equation \ref{eq:qc-state} for the last inequality.  Since the $E$ and $A_Z$ registers are independent in $\chi^{(y)}$ we have $\Hmin(A_Z|E)_{\chi^{(y)}} = \Hmin(A_Z)_{\chi^{(y)}} = -\log_2\max_zp(z|y)$.  It is not difficult to see that $p(z|y) = d^{-n}$ for all $y,z \in \al_d^n$.  Thus $\Hmin(A_Z|E)_{\chi} \ge n\log_2 d$.  Note that our bound here, and also on $|J(q\st x)|$, are independent of $x$.  Thus, concluding, we have the following bound on the entropy in the ideal state:
\begin{equation}\label{eq:ideal-entropy}
  \Hmin(A_Z|E)_{\sigma} \ge \min_x\Hmin(A_Z|E)_{\sigma^{(x)}} \ge n\left(\log_2 d - \frac{H_d(w(s(q)) + \delta)}{\log_d 2}\right).
\end{equation}
Of course, this was only the ideal state analysis, however, Equation \ref{eq:ideal-entropy} holds for any choice of subset $t$ and observation $q$.  We now use this result to derive the final security of the real state produced by Eve and show that, with high probability over the choice of subset $t$ and measurement outcome $q$, the final secret key produced by the protocol will be secure.

\textbf{Step 3: Real State Security -} The QCKA protocol (and, indeed, most if not all QKD protocols) may be broken into three distinct modules or CPTP maps: first is a sampling module $\mathcal{S}$ which takes as input a quantum state $\rho_{TABE}$ where the $T$ register represents the sampling subset $t$ used and $B$ represents all $p$ Bobs.  Here, this module measures the $T$ register which chooses a subset $t$; from this, all qudits indexed by $t$ are measured in the Fourier basis, producing outcome $q \in \al_d^{m\cdot(p+1)}$.  The output of this process is the subset chosen $t$, the observed $q$, and also the post-measured state $\rho_{ABE}(t,q)$.  Following this, the raw-key generation module is run, denoted $\mathcal{R}$, which takes as input the previous post measured state and measures the remaining systems in the $Z$ basis resulting in raw keys for all parties.  The output of this module is the raw key produced along with a post-measured state for Eve.  Finally, a post-processing module is run, denoted $\mathcal{P}$, which will run an error correction protocol and privacy amplification, yielding the final secret key.  The output of this last CPTP map is the actual secret key produced along with Eve's final quantum ancilla.  This module requires as input the raw keys along with $q$ (needed to determine the final secret key size).  We want to show, with high probability over the choice of sampling subset and test measurement outcome, that the final secret key is $\epsilon_{PA}$-close to the ideal secret key as defined by Equation \ref{eq:PA}.

Recall, $\ket{\psi}_{AB_1\cdots B_pE}$ is the actual state produced by the adversary and sent to each of the parties.  We may assume this is a pure state as a mixed state would lead to greater uncertainty for Eve.  Of course, in the real case, the choice of subset is independent of the state produced by Eve and so we write the complete real state as $\rho_{TABE} = \sum_t\frac{1}{T}\kb{t}\otimes\kb{\psi}$ where $T = {N \choose m}$.  From this, an ideal state of the form $\sum_t\frac{1}{T}\kb{t}\otimes\kb{\phi^t}_{ABE}$ may be defined as was analyzed previously in the second step of the proof.  We may write the action of the composition $\mathcal{P}\circ\mathcal{R}\circ\mathcal{S} = \mathcal{PRS}$ as follows:
\begin{align}
  \mathcal{PRS}\left(\sum_t\frac{1}{T}\kb{t}_T\otimes\kb{\psi}\right) &= \sum_{q,t}p(q, t) \kb{q,t} \otimes\mathcal{P}_q\mathcal{R}\left(\kb{\psi_q^t}_{ABE}\right)\\
  \mathcal{PRS}\left(\sum_t\frac{1}{T}\kb{t}_T\otimes\kb{\phi^t}\right) &= \sum_{q,t}\tilde{p}(q, t) \kb{q,t} \otimes\mathcal{P}_q\mathcal{R}\left(\kb{\phi_q^t}_{ABE}\right).
\end{align}
Above, $p(q,t)$ is the probability of choosing subset $t$ and observing outcome $q$ in the real state and $\tilde{p}(q,t)$ is similar but for the ideal state.  The post-measured state after sampling are denoted $\ket{\psi_q^t}$ in the real case and $\ket{\phi_q^t}$ in the ideal case (see Equation \ref{eq:ideal-post-measure} for what this state looks like in the ideal case).  Note that, conditioning on a particular $q$ and $t$, these states are pure.

Let $\ell(q,\leak) = n(\log_2 d - \frac{1}{\log_d 2}H_d(w(s(q)) + \delta)) - \leak - 2\log_2\left(\frac{1}{\epsilon}\right)$ where $\leak$ will be used to denote the leaked information due to error correction.  Then, from Equation \ref{eq:PA} and our analysis on the min entropy of the post-measured ideal state in Equation \ref{eq:ideal-entropy}, we know that for any $t$ and observed $q$, if privacy amplification shrinks the raw key to a size of $\ell$, it holds that:
\begin{equation}
  \trd{\mathcal{P}_q\mathcal{R}\left(\kb{\phi_q^t}\right) - \mathcal{U}_{\ell(q,\leak)}\otimes tr_A \mathcal{P}_q\mathcal{R}\left(\kb{\phi_q^t}\right)} \le \epsilon,
\end{equation}
where $\mathcal{U}_k = \frac{1}{2^k}\sum_{i=0}^{2^k-1}\kb{i}$ is an operator acting on $\mathcal{H}_{2^{n\log_2 d}}$ (note that $n\log_2 d$ is the largest number of bits the final secret key can possibly be; privacy amplification will hash this into something potentially smaller and so $\mathcal{U}$ represents a uniform distribution on this smaller subspace of potential secret keys).  Note that, above and in the text below, we are tracing out the $B$ systems though we do not explicitly write out $tr_B$ in all equations as it would add additional, and unnecessary, bulk to the equations.  Hence, from here on out, the reader may assume all Bob systems are traced out of the equations unless otherwise stated.  Finally, note that the above of course implies that:
\begin{equation}\label{eq:ideal-epPA}
    \trd{\sum_{q,t}\tilde{p}(q,t)\kb{q,t}\otimes\mathcal{P}_q\mathcal{R}\left(\kb{\phi_q^t}\right) - \sum_{q,t}\tilde{p}(q,t)\kb{q,t}\otimes\mathcal{U}_{\ell(q,\leak)}\otimes tr_A \mathcal{P}_q\mathcal{R}\left(\kb{\phi_q^t}\right)} \le \epsilon,
\end{equation}

We now claim that, with high probability over $t$ and measurement outcome $q$, it holds that:
\begin{equation}\label{eq:pa-claim}
\trd{\mathcal{P}_q\mathcal{R}\left(\kb{\psi_q^t}\right) - \mathcal{U}_{\ell(q,\leak)}\otimes tr_A\mathcal{P}_q\mathcal{R}\left(\kb{\psi_q^t}\right)} \le 5\epsilon + \left(20\epsilon\right)^{1/3} = \epsilon_{PA}
\end{equation}
thus ensuring, again with high probability over the subset choice and test measurement outcome, that the resulting secret key in the real case, using the state produced by the adversary, is $\epsilon_{PA}$ close to an ideal secret key.

Let $\rho_{TABE}$ and $\sigma_{TABE}$ be the real and ideal states respectively.  Now, since the ideal and real states are $\epsilon$-close in trace distance by Theorem \ref{thm:sample}, along with our choice of $\delta$ and our sampling strategy, and since quantum operations cannot increase trace distance, we have:
\begin{align}
  2\epsilon &\ge \trd{\rho - \sigma} \ge \trd{\mathcal{P}\mathcal{R}\mathcal{S}(\rho) - \mathcal{P}\mathcal{R}\mathcal{S}(\sigma)}\notag\\
  &=\trd{\sum_{q,t}p(q,t)\kb{q,t}\otimes\mathcal{P}_q\mathcal{R}(\kb{\psi_q^t}) - \sum_{q,t}\tilde{p}(q,t)\kb{q,t}\otimes\mathcal{P}_q\mathcal{R}(\kb{\phi_q^t})}\label{eq:diff-ideal-real}
\end{align}
From the above, we have:
\begin{equation}\label{eq:diff-U}
  2\epsilon \ge \trd{\sum_{q,t}p(q,t)\kb{q,t}\otimes\mathcal{U}_{\ell(q,\leak)}\otimes tr_A\mathcal{P}_q\mathcal{R}(\kb{\psi_{q}^t}) - \sum_{q,t}\tilde{p}(q,t)\kb{q,t}\otimes\mathcal{U}_{\ell(q,\leak)}\otimes tr_A\mathcal{P}_q\mathcal{R}(\kb{\phi_{q}^t})}
\end{equation}
This follows from basic properties of trace distance along with the fact that partial trace is a quantum operation.

Adding the Equations \ref{eq:diff-ideal-real} and \ref{eq:diff-U} above yields:
\begin{align*}
  4\epsilon &\ge \trd{\sum_{q,t}p(q,t)\kb{q,t}\otimes\mathcal{P}_q\mathcal{R}(\kb{\psi_q^t}) - \sum_{q,t}\tilde{p}(q,t)\kb{q,t}\otimes\mathcal{P}_q\mathcal{R}(\kb{\phi_q^t})}\\
  &+ \trd{\sum_{q,t}p(q,t)\kb{q,t}\otimes\mathcal{U}_{\ell(q,\leak)}\otimes tr_A\mathcal{P}_q\mathcal{R}(\kb{\psi_{q}^t}) - \sum_{q,t}\tilde{p}(q,t)\kb{q,t}\otimes\mathcal{U}_{\ell(q,\leak)}\otimes tr_A\mathcal{P}_q\mathcal{R}(\kb{\phi_{q}^t})}\\
  &\ge \trd{\sum_{q,t}p(q,t)\kb{q,t}\otimes\left(\mathcal{P}_q\mathcal{R}(\kb{\psi_q^t}) - \mathcal{U}_{\ell(q,\leak)}\otimes tr_A\mathcal{P}_q\mathcal{R}(\kb{\psi_q^t})\right)}\\
  & - \trd{\sum_{q,t}\tilde{p}(q,t)\kb{q,t}\otimes\left(\mathcal{P}_q\mathcal{R}(\kb{\phi_q^t}) - \mathcal{U}_{\ell(q,\leak)}\otimes tr_A\mathcal{P}_q\mathcal{R}(\kb{\phi_q^t})\right)}\\&\ge \trd{\sum_{q,t}p(q,t)\kb{q,t}\otimes\left(\mathcal{P}_q\mathcal{R}(\kb{\psi_q^t}) - \mathcal{U}_{\ell(q,\leak)}\otimes tr_A\mathcal{P}_q\mathcal{R}(\kb{\psi_q^t})\right)} - \epsilon,
\end{align*}
where, above we used the triangle inequality followed by the reverse triangle inequality and finally Equation \ref{eq:ideal-epPA}.  Let $\Delta_{t,q} = \frac{1}{2}\trd{\mathcal{P}_q\mathcal{R}(\kb{\psi_q^t}) - \mathcal{U}_{\ell(q,\leak)}\otimes tr_A\mathcal{P}_q\mathcal{R}(\kb{\psi_q^t})}$.  Then, the above, along with basic properties of trace distance, implies:
\[
\frac{5\epsilon}{2} \ge \sum_{q,t}p(q,t)\Delta_{q,t}.
\]
We now consider $\Delta_{q,t}$ as a random variable over $q$ and $t$.  From the above, its expected value is upper-bounded by $5\epsilon/2$.  Furthermore, since $\Delta_{t,q} \le 1$ for all $t,q$ (by properties of trace distance), the variance may also be upper-bounded by $5\epsilon/2$.  Using Chebyshev's inequality, then, we have:
\begin{equation}
  Pr\left[\left|\Delta_{t,q} - \frac{5\epsilon}{2}\right| \le \left(\frac{5\epsilon}{2}\right)^{1/3}\right] \ge 1 - \left(\frac{5\epsilon}{2}\right)^{1/3},
\end{equation}
From this, and simple algebra, it follows that, except with probability at most $\epsilon_{\text{fail}} = (5\epsilon/2)^{1/3}$, Equation \ref{eq:pa-claim} holds.  This implies that, with high probability over the choice of subset $t$ and test measurement outcome in the Fourier basis $q$, Alice and the $p$ Bob's are left with an $\epsilon_{PA} = 5\epsilon + (20\epsilon)^{1/3}$ secure key of size:
\begin{equation}\label{eq:final-keyrate}
  \ell = n\left(\log_2d - \frac{H_d(w(s(q)) + \delta)}{\log_d2}\right) - \leak - 2\log_2\frac{1}{\epsilon}.
\end{equation}
concluding the security proof.

\subsection{Evaluation}

We now evaluate our key-rate bound for this protocol.  We will first consider the two-dimensional case, allowing us to compare with current state of the art results from \cite{finite-ghz-bb84}.  We will then evaluate our bound in higher dimensions - in that case, we have no other QCKA results to compare to (the results in \cite{finite-ghz-bb84} applied only to the qubit case); however, we will show some interesting behavior in the higher-dimensional case, when compared to the qubit case. 

To evaluate, we will assume a depolarization channel connecting all parties.  This assumption is not required for our security proof which works for any channel - one must simply observe the value $q$ and also the error correction leakage used by the EC protocol and then evaluate the secret key rate (Equation \ref{eq:final-keyrate}) using our analysis in the prior section.  However, we will consider depolarization channels in this subsection in order to evaluate our bound here without actual hardware, and also to compare with prior work (which also assume depolarization channels when evaluating key-rates).

Under a depolarization channel, we may assume the quantum state shared by Alice and the $p$ Bobs is of the form:
\begin{equation}\label{eq:dep-result}
\rho_{AB}^{\otimes N} = \left( (1-Q)\kb{\psi_0} + \frac{Q}{d^{p+1} }I\right)^{\otimes N},
\end{equation}
where $I$ is the identity operator of dimension $d^{p+1}$.  Note that, under this assumption, the expected value of $w(s(q))$ is $Q/d$.  This matches the value evaluated in \cite{finite-ghz-bb84} for the qubit case as expected (where, there, the $X$ basis was used and a parity check performed).

We next need a bound on $\leak$.  In practice, this can be done through the actual public transcript after executing the protocol; however for our evaluation, we will simulate an expected leakage.  For error correction (EC), we assume one-way error correction and take the same approach as in \cite{finite-ghz-bb84}, whereby Alice will send the same error correction information to each of the $p$ Bob's.  In particular, it was proven there, that there exists a one-way EC protocol for such a scenario that aborts with probability no greater than $2p\epsilon'$ where the leakage is upper-bounded by:
\[
\leak \le \max_iH_0^{\epsilon'}(A|B_i) + \log_2\frac{2(N-1)}{\epsilon_{EC}}
\]
where:
\[
(1-2p\epsilon')Pr(\exists i \st B_i \ne A \text{ after EC}) \le \epsilon_{EC}
\]
and where $H_0^{\epsilon'}(A|B_i)$ is the smooth R\'enyi zero-entropy of Alice's raw key conditioned on the $i$'th Bob's, namely:
\[
H_0^{\epsilon'}(X|Y) = \min_{P_{XY}}\max_y\text{supp}(P(X|Y=y))
\]
where the minimum is over all probability distributions $P$ that are $\epsilon'$-close to the original input distribution.  Importantly, one need only consider the ``worst-case'' noise between Alice and one Bob, as opposed to taking the sum of all error correction leakages for all $p$ Bob's.  A single error correction message from Alice is sufficient to correct all $p$ Bob's raw keys.  To ensure error correction succeeded, Alice will choose a random two-universal hash function $f$, the output size of which we take to be $\log_2\frac{1}{\epsilon_{EC}}$-bits, and broadcast $f(A)$, where $A$ is her raw key as discussed earlier when introducing the protocol.  This leaks an additional $\log_2\frac{1}{\epsilon_{EC}}$ bits which must be deducted from the final secret key size.  This is used so users can be assured that error correction has succeeded.

Using results from \cite{HD-BB84} to bound the R\'enyi zero-entropy in this high-dimensional scenario, along with the depolarization assumption, we may bound the error correction leakage by:
\[
\leak \le nh(Q_Z + \nu) + n(Q_Z+\nu)\log_2(d-1) + \log_2\frac{1}{\epsilon_{EC}},
\]
where:
\[
\nu = \sqrt{\frac{N(m+1)\ln\frac{4p}{\epsilon_{EC}}}{m^2(N-m)}}.
\]
and where $Q_Z = \max_i Q_i$, where $Q_i$ is the probability of an error in Alice and the $i$'th Bob's raw key digit.  Note that we are using the same sample size $m$ used for the Fourier basis measurement test and this must be deducted from the total raw key size.  Since we are evaluating assuming a depolarization channel, we have $Q_Z = Q(1-1/d)$ (which is easily seen from Equation \ref{eq:dep-result}).  Note that, we use this only for evaluation purposes as it will allow us to directly compare, in the qubit case, to state-of-the-art results in \cite{finite-ghz-bb84}.

Combining everything, we find the length of the key produced by the protocol to be:
\begin{equation}
  \ell = n\left(\log_2 d - \frac{H_d\left(\frac{Q}{d} + \delta\right)}{\log_d 2} - h\left(Q_Z + \nu\right) - (Q_Z+\nu)\log_2(d-1)\right) - \log_2\frac{1}{\epsilon_{EC}} - 2\log_2\frac{1}{\epsilon}.
\end{equation}
Of course the actual key-rate, then, is simply $\ell / (n+2m)$ (we divide by an additional $m$ number of signals to account for the sampling of the raw-key needed to estimate $Q_Z$ above).  In our evaluations, we set $\epsilon_{EC} = 10^{-12}$ and $\epsilon = 10^{-36}$ giving a failure probability (both for the entropy bound and error correction) on the order of $10^{-12}$.  This also sets $\epsilon_{PA}$ to be on the order of $10^{-12}$.  When comparing with other protocols, we use a failure probability of $10^{-12}$.  Finally, we use a sample size of $7\%$ for both bases (i.e., $m = .07N$ where $N$ is the total number of signals sent).

\begin{figure}
    \centering
    \includegraphics[width=.7\textwidth]{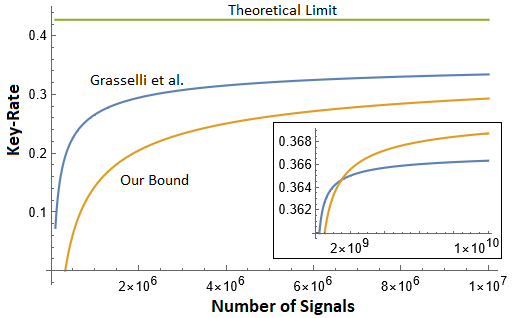}
    \caption{Comparing our new bound with that from \cite{finite-ghz-bb84} for the qubit case ($d=2$) when $Q = 10\%$.  We note that our result is slightly lower than in \cite{finite-ghz-bb84} for this dimension.  However, the advantage to our approach is that it can readily handle higher dimensions.  Inset: a close-up view of the difference in our bound and that from \cite{finite-ghz-bb84}; we note that as the number of signals increases, our results converge.  See text for discussion.}
    \label{fig:comp1}
\end{figure}

A comparison of our key-rate bound, and that derived in \cite{finite-ghz-bb84} through alternative means, is shown in Figure \ref{fig:comp1} for the two dimensional case (in which case both protocols are identical).  We note that, except for a slight deviation, the two results agree (with prior results from \cite{finite-ghz-bb84} surpassing ours by a small amount).  Of course, the proof and results in \cite{finite-ghz-bb84} apply only to $d=2$; to our knowledge, we are the first to derive a rigorous finite-key proof of security for a high-dimensional QCKA protocol.

\begin{figure}
    \centering
    \includegraphics[width=.7\textwidth]{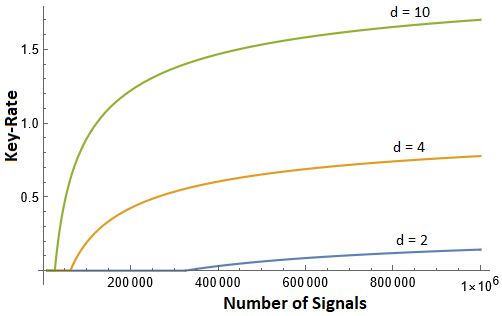}
    \includegraphics[width=.7\textwidth]{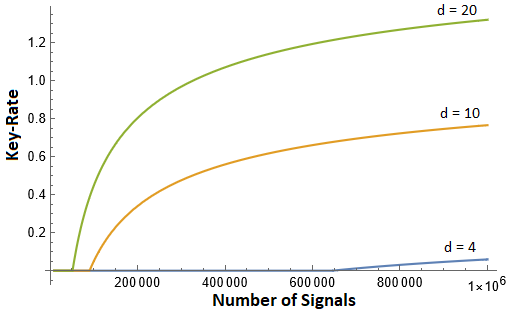}
    \caption{Evaluating our key-rate bound for higher dimensions assuming $Q = 10\%$ (Top) and $Q = 30\%$ (Bottom).  We note that, as dimension increases, not only does key-rate increase, but also noise tolerance.  Furthermore, the number of signals required before a positive key-rate is attained also decreases with dimension for a fixed noise level.}
    \label{fig:hd-keyrate}
\end{figure}

\begin{figure}
    \centering
    \includegraphics[width=.7\textwidth]{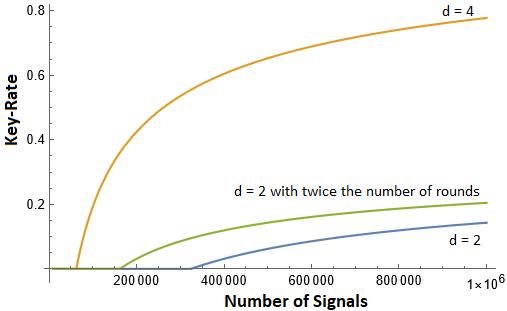}
    \caption{Showing that the advantage in key-rate for higher dimensions cannot be recovered simply by using lower-dimensional systems and increasing the number of rounds.  High-dimensional states exhibit an advantage beyond simple parallel executions of a qubit-based protocol for this QCKA protocol.}
    \label{fig:comp3}
\end{figure}

We also evaluate our key-rate bound in higher-dimensions in Figure \ref{fig:hd-keyrate}.  In higher dimensions, we cannot compare to any other QCKA protocols as we are not aware of any other finite key security results for such protocols in high (greater than $2$) dimensions.    However, we note several interesting properties here.  First, as the dimension increases, the number of signals needed before a positive key-rate is achieved, decreases, and the general key-rate increases,  making the protocol potentially more efficient.  Note that one explanation for the increased key-rate is due to the fact that one receives, for each signal, a larger number of raw-key bits as the dimension increases.  However this, alone, does not explain the great increase in key-rate as the signal dimension increases.  For instance, if we compare $d=2$ and $d'=4$, a single iteration of the protocol, in the first case, produces at most one raw key bit, while the second case would produce at most $2$ raw key bits.  If this were the only reason for the increase in secret key-rates, one would expect that running twice the number of iterations for the $d=2$ case would produce the same secret key length as the $d'=4$ case.  However this is clearly not the case, as shown in Figure \ref{fig:comp3}.  Thus, the increase in key-rate for higher dimensions cannot be recovered simply by running multiple copies of the qubit-based protocol in parallel, instead higher-dimensional states per round are required.  We also note that the number of Bob's, $p$, does not noticeably affect the key-rate - interestingly, this was also discovered in \cite{finite-ghz-bb84} for the qubit, $d=2$, case.

\section{Closing Remarks}

In this paper, we proved the security of a high-dimensional QCKA protocol, allowing multiple parties to establish a shared secret key.  We proved security using a combination of the quantum sampling framework of \cite{sample}, along with sampling-based entropic uncertainty relation techniques from \cite{krawec2019quantum}.  Our proof introduced several new methods needed to use those two frameworks in this multi-user scenario and our methods may be applicable to other multi-user quantum cryptographic protocols, especially in higher dimensions.  Finally, we evaluated the protocol in a variety of scenarios and showed some interesting properties in higher-dimensions.  Our work here has shown even more evidence, beyond that already known (as discussed in the Introduction), of the potential benefits, at least in theory, of high-dimensional quantum states.  Note that we did not consider practical device imperfections, leaving that as interesting future work.

$ $\newline\newline
\footnotesize{\textbf{Acknowledgments: } WOK would like to acknowledge support from the National Science Foundation under grant number 2006126.}
$ $\newline\newline
\footnotesize{\textbf{Disclaimer: }This paper was prepared for information purposes by the teams of researchers from the various institutions
identified above, including the Future Lab for Applied Research and Engineering (FLARE) group of JPMorgan
Chase Bank, N.A.. This paper is not a product of the Research Department of JPMorgan Chase \& Co. or its
affiliates. Neither JPMorgan Chase \& Co. nor any of its affiliates make any explicit or implied representation
or warranty and none of them accept any liability in connection with this paper, including, but limited to,
the completeness, accuracy, reliability of information contained herein and the potential legal, compliance, tax
or accounting effects thereof. This document is not intended as investment research or investment advice, or
a recommendation, offer or solicitation for the purchase or sale of any security, financial instrument, financial
product or service, or to be used in any way for evaluating the merits of participating in any transaction.}


\end{document}